\documentclass[12pt]{article}
\pdfoutput=1
\usepackage{subfigure}
\usepackage{amssymb,amsmath}
\usepackage{graphicx}
\usepackage{color}
\usepackage[colorlinks=true
,urlcolor=blue
,citecolor=blue
,linkcolor=blue
,pagecolor=blue
,linktocpage=true
,pdfproducer=medialab
]{hyperref}
\usepackage[a4paper,width=16.8cm]{geometry}
\makeatletter \renewcommand{\@dotsep}{10000} \makeatother
\usepackage{appendix}

\begin{document}

\begin{center}

 {\Large\bf A Hamiltonian for Massless and Zero-Energy States 
 } \vspace{1cm}

{   Muhammad Adeel Ajaib\footnote{ E-mail: adeel@udel.edu}}

{\baselineskip 20pt \it
Department of Physics, Coastal Carolina University, Conway, SC, 29526 \\
Department of Mathematics, Statistics and Physics, Qatar  University,  Doha,  Qatar
 } \vspace{.5cm}

{\baselineskip 20pt \it
   } \vspace{.5cm}

\setcounter{footnote}{0}
\vspace{1.5cm}
\end{center}

\begin{abstract}

{We present a non-hermitian Hamiltonian which can be employed to explain a condensed matter system with effectively massless and zero energy states.} We analyze the 2D tunneling problem and derive the transmission and reflection coefficients for the massless and zero energy states.  We find that the transmission coefficient of the massless and zero-energy particles in this case is consistent with non-chiral tunneling of quasiparticles which is in contrast to the Klein tunneling known for electrons described by the Dirac equation. Our analysis predicts that the massless electron can be reflected as a hole-like state whereas this transition does not occur for the zero-energy state. 
Experimental observations are needed to test whether the presented Hamiltonian can be realized in such a condensed matter system.

\end{abstract}

\newpage

\section{Introduction}\label{intro}

Graphene is one of the most important materials in contemporary
condensed matter physics due to the very interesting properties it possesses. It is essentially a 2D system that consists of a single layer of carbon atoms arranged in a honey comb lattice. Due to graphene’s crystal symmetry, quasi-particles near Dirac points behave as massless Dirac fermions (for a review see \cite{review}).

 The massless electrons in graphene are typically described by employing the massless Dirac equation which respects Lorentz invariance. In this paper we present a Hamiltonian that results from Lorentz violating terms and yields the dispersion relation for massless electron/hole-like and zero-energy states. This Hamiltonian can be relevant to graphene like condensed matter systems where massless and zero-energy states are known to exist in various scenarios. Zero energy states, which typically arise as bound states in graphene, have attracted considerable interests in recent years \cite{zero1}. {These states, for example, can arise in graphene in the presence of a magnetic field, as the lowest Landau level state \cite{zeromag} and in the presence of electric fields as well \cite{zeroelec}}. These states have also been described in the context of supersymmteric quantum mechanics \cite{Halberg:2017lim} and also for various potentials \cite{zero1}. In addition, zero energy states can also arise as edge or surface states in topological insulators \cite{hasan:2010}.
 
The paper is organized as follows: In section \ref{sec:hamiltonian}, we present the Hamiltonian and in section \ref{sec:2dtunnel}, we solve the problem of 2D tunneling of electrons for a potential barrier. In section \ref{sec:2dtunnel}, we also solve for the transmission and reflection coefficients of the massless and zero-energy electrons. We conclude in section \ref{conclude}.

\section{The Hamiltonian }\label{sec:hamiltonian}

In this section, we present a {non-hermitian} Hamiltonian which can be employed to describe the zero-energy and massless states that arise in {graphene-like} condensed matter systems. {There is a growing interest in non-hermitian Hamiltonians that can be employed, for instance, to study topological edge states \cite{Yuce:2018gtt}.}
{The Hamiltonian in $2+1$ dimensional space-time is given by ($\hbar=c=1$)}
\begin{eqnarray}
H= \eta(\gamma_i p_i+m)
\label{eq:h1}
\end{eqnarray}
where $\gamma_i=(\gamma_1,\gamma_2)$ and $\eta=\frac{1}{2}\gamma_5(I-\gamma_0)$ is a non-symmetric nilpotent matrix ($\eta^2=0$). {The gamma matrices employed herein are the usual $4 \times 4$ Dirac matrices.} Due to the presence of the mass term, in particular, the $\gamma_5 \gamma_0 m$ term,  the Hamiltonian is not hermitian. The energy eigenvalues of the above Hamiltonian are two degenerate zero-energy electron/hole-like states ($E=0$) and the other two represent massless electron/hole-like states ($E=\pm \sqrt{p_x^2+p_y^2}$). The Dirac Hamiltonian for massless quasiparticles in graphene results from its crystal symmetry. In the presence of impurities, however, the Hamiltonian can be modified \cite{Ardenghi} and it may be possible to realize a system consistent with the dynamics of the above Hamiltonian. {Experimental observations are needed to test whether the above Hamiltonian might be realized in a Graphene-like system}.

Notice that even with the presence of a mass term in the Hamiltonian (\ref{eq:h1}) the dispersion relation describes particles that are effectively massless and have zero energy. Following the procedure similar to \cite{Ajaib:2015uha}, we can find the current densities corresponding to the above Hamiltonian as
\begin{eqnarray}
J_i &=& \psi^\dagger (\Gamma \gamma_i \eta^\dagger \eta) \psi \\
\rho &=&  \psi^\dagger \gamma_5 \psi 
\end{eqnarray}
where, $\Gamma=\gamma_3 \gamma_5 \gamma_0$. Therefore, the current density is not hermitian whereas the probability density is hermitian and positive definite. Note that this current density can only be obtained in the one or two dimensional case.

It has been proposed that several condensed matter phenomena can be described with enhanced Lorentz violation operators in the context of the Standard Model Extension (SME) \cite{Ajaib:2012wq}.  The Hamiltonian (\ref{eq:h1}) includes terms that violate Lorentz invariance and these terms, except for the $\gamma_5 \gamma_0 m$ term, can be found in  \cite{Kostelecky:1999zh} (see equation (7) of this reference). The term $\gamma_5 \gamma_0 m$, however,  is non-hermitian and is not part of the SME.

%\newpage
\begin{figure}
\vspace*{-1.2cm}
\centering
\includegraphics[scale=.39]{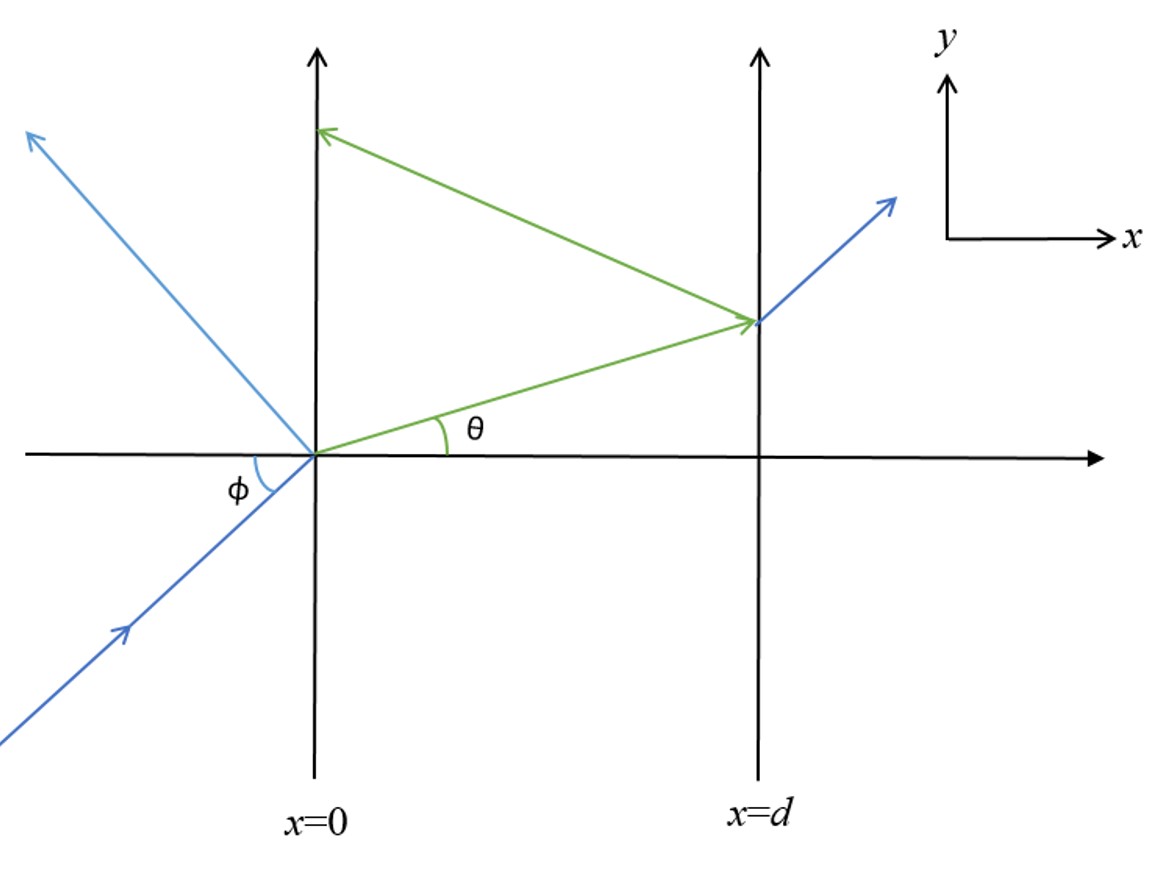}
\includegraphics[scale=.35]{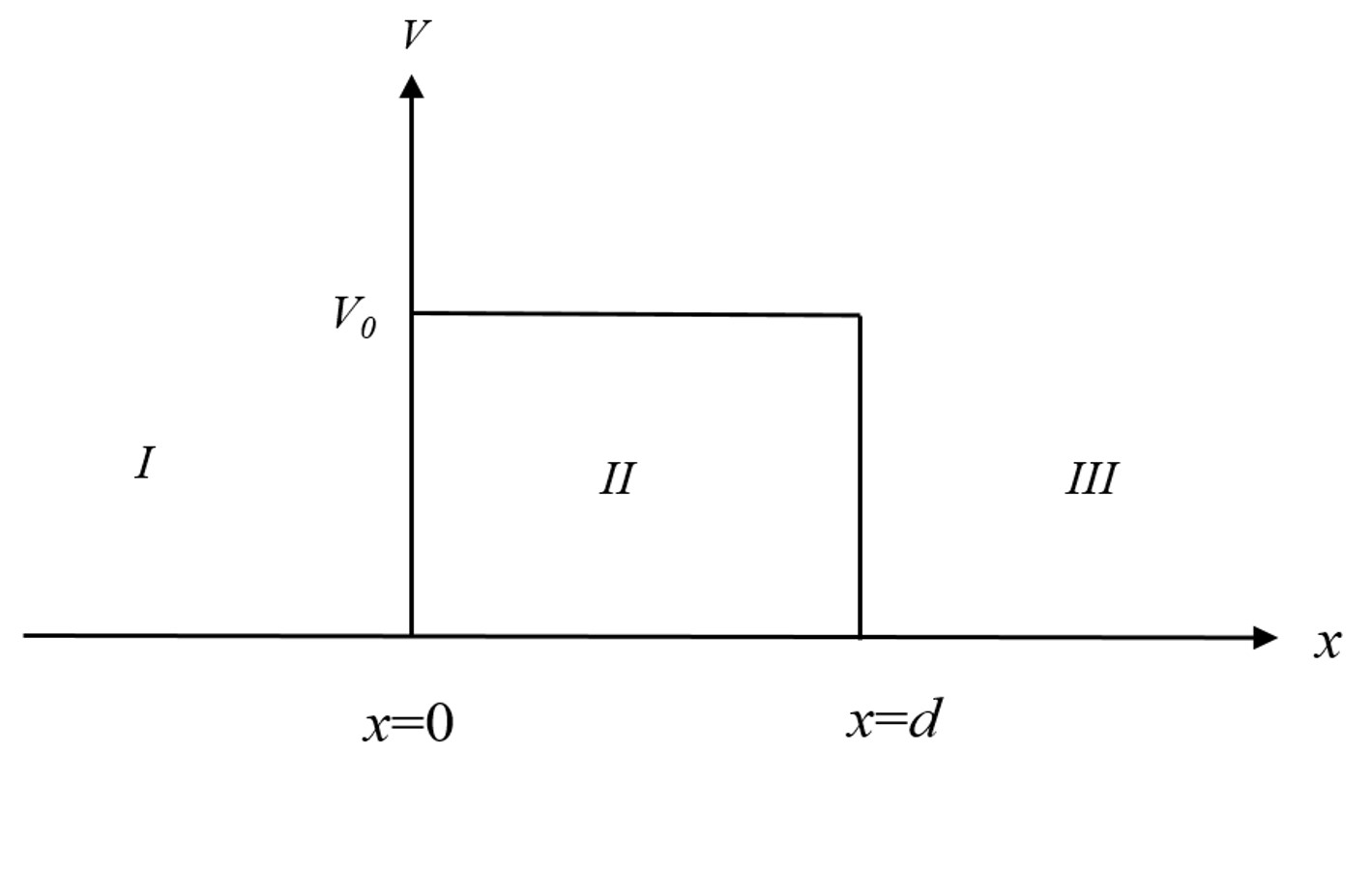}
\caption{Schematic representation of the 2D barrier tunneling problem.}
\label{fig:2dbarrier}
\end{figure}

\section{Barrier Tunneling in 2D } \label{sec:2dtunnel}

\subsection{Massless States}

In this section, we analyze the 2D scattering of an electron incident on a potential barrier of height $V_0$ and width $d$ (schematically shown in Figure \ref{fig:2dbarrier}).  We analyze the case of an electron with energy $E$ incident on a potential barrier of height $V_0$ and width $d$.  
\begin{eqnarray}
H= v \eta(\hbar \gamma_i k_i+m v)
\label{h1}
\end{eqnarray}
where we have replaced the velocity of light by  the Fermi velocity  $ {\it v}$ of electrons in graphene, $ {\it v}\approx 10^6 m/s$. 
 The eigenstates corresponding to the massless electron/hole-like states are given by 
\begin{eqnarray}
u(k_x,k_y) = 
\left(
\begin{array}{cc}
 \frac{-k_x+i k_y}{\sqrt{k_x^2+k_y^2}}   \\
 1   \\
 0 \\
0
\end{array}
\right), \ \ \ \
\label{matrix:eta2d}
%\end{eqnarray}
%
%\begin{eqnarray}
%
v(k_x,k_y) = 
\left(
\begin{array}{cc}
 \frac{k_x-i k_y}{\sqrt{k_x^2+k_y^2}}\\
1\\
 0   \\
 0   
\end{array}
\right).
\label{eigenstates:e1}
\end{eqnarray}
\begin{figure}
\centering
\includegraphics[scale=.46]{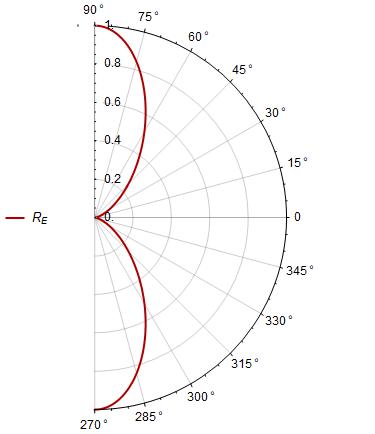}
\includegraphics[scale=.46]{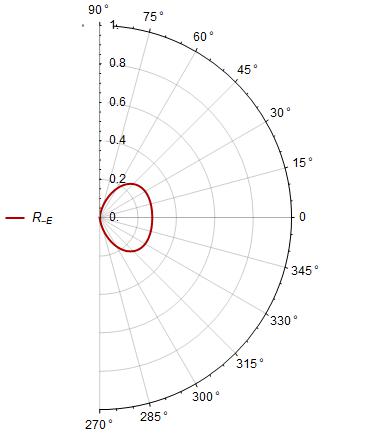}
\caption{The plot shows the reflection coefficients ($R_{\pm E}=|r_{\pm E}|^2$) for an electron as a function of the incident angle $\phi$, for $E>V_0$, given in equations (\ref{eq:rfcoeffs}). The left panel shows the probability that the electron is reflected as an electron whereas the panel on the right shows the probability of the electron to be reflected as a hole. The energy of the incident electrons, the height and width of the potential barrier are chosen to be $E=80$ meV, $V_0=$70 meV and d=10 nm.}
\label{fig:ref-coeff}
\end{figure}

\begin{figure}
\centering
\includegraphics[scale=.47]{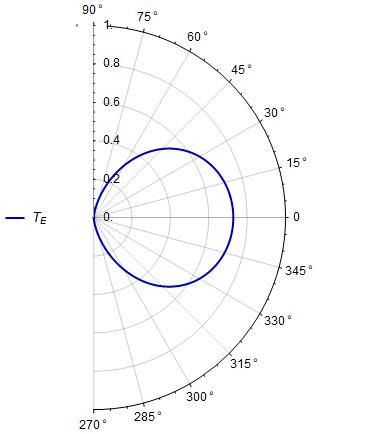}
\caption{The plot shows the transmission coefficient  ($T_{ E}=|t_{ E}|^2$)  for electrons as a function of the incident angle $\phi$, for $E>V_0$, given in equations (\ref{eq:transmission}). The analysis predicts that the electron will always be transmitted as an electron. The energy of the incident electrons, the height and width of the potential barrier are chosen to be $E=80$ meV, $V_0=$70 meV and d=10 nm.}
\label{fig:tran-coeff}
\end{figure}
The wave functions in the three regions shown in Figure \ref{fig:2dbarrier} are given by
\begin{eqnarray}
\psi_{I}(x, y) &=&e^{i k_y y}  (u(k_x,k_y) e^{i k_x x} +r_{+ E} \ u(-k_x,k_y) e^{-i k_x x} +r_{- E} \ v(-k_x,k_y) e^{-i k_x x}) \nonumber\\[10pt]
\psi_{II}(x, y) &=& e^{i k_y y}  (a_1   u(q_x,k_y) e^{i q_x x} + a_2   u(q_x,k_y) e^{i q_x x} + b_1  u(-q_x,k_y) e^{-i q_x x} +b_2  u(-q_x,k_y) e^{-i q_x x} )  \nonumber \\[10pt]
\psi_{III}(x, y) &=& e^{i k_y y} (t_{+ E} \ u(k_x,k_y) e^{i k_x x}  + t_{- E} \ v(k_x,k_y) e^{i k_x x} )
\label{eq:wave-fn}
\end{eqnarray}

\vspace{3mm}
\noindent Here, $k_y=k \sin\phi$, $k_x=k \cos\phi$,  $q_x=q \cos\theta$. The conservation of the wave vector in the $y$-direction implies $k \sin\phi=q \sin\theta$. Applying the continuity of the wave function and its derivative at the boundaries $x=0$ and $x=d$ yields the following transmission coefficient for the effectively massless electron is given by
\begin{eqnarray}
t_{+ E}=\frac{4 e^{-i k_x d}k_x q_x}{(q_x+k_x)^2 e^{-i k_x d}-(q_x - k_x)^2 e^{i k_x d}}
\label{eq:transmission}
\end{eqnarray}
whereas the transmission coefficient corresponding to the probability of the electron to be transmitted as a hole is zero, $t_{-E}=0$. The transmission coefficient  (\ref{eq:transmission})  corresponds to non-chiral electrons \cite{Katsnelson:2006} and implies that the barrier is transparent under resonance conditions $q_x d=n\pi$ ($n=0,\pm1,..$) whereas the coefficient is a function of the tunneling parameters at normal incidence ($\phi=0$).  This behavior is in contrast to chiral electrons described by the Dirac equation which exhibits Klein tunneling and the barrier is transparent at normal incidence  \cite{Katsnelson:2006}.

The reflection coefficients for the effectively massless electron ($r_{+E}$) and hole ($r_{-E}$) are given by
\begin{eqnarray}
r_{+E} &=&\frac{1}{k_x+i k_y}\  \frac{2 k_y (k_x - q_x) (k_x + q_x) \sin(d q_x)}{(q_x+k_x)^2 e^{-i k_x d}-(q_x - k_x)^2 e^{i k_x d}} \nonumber \\
r_{-E} &=& \frac{-i}{k_x+i k_y}\  \frac{2 k_x (k_x - q_x) (k_x + q_x) \sin(d q_x)}{(q_x+k_x)^2 e^{-i k_x d}-(q_x - k_x)^2 e^{i k_x d}}
\label{eq:rfcoeffs}
\end{eqnarray}
with $|t_E|^2+|r_{+E}|^2+|r_{-E}|^2=1$. In Figures \ref{fig:ref-coeff} and \ref{fig:tran-coeff}, we display the reflection and transmission coefficients as a function of the incident angle $\phi$ for the case when the incident electron  energy is $E=80$ meV, the barrier height is $V_0=$70 meV and barrier width is $d$=10 nm. We can see from the figures that for relatively small angles ($\lesssim 45^{\circ}$) the probability of the electron to be transmitted is fairly significant, whereas, there is a small probability for the electron to be reflected as a hole{, i.e., the electron jumps from the conduction to the valence band}. For relatively large angles ($\gtrsim 60^{\circ}$), we can see from the left panel of Figure  \ref{fig:ref-coeff} that electron is most likely to be reflected as an electron. For normal incidence ($\phi \simeq 0$ or $k_y \simeq 0$ ), the coefficients predict that for small angles the electron is most likely to be reflected as a hole.

%Lastly, we address the question of whether the electron or the hole can transit to a zero energy state. We have examined this case and this is not a possible outcome of the analysis.

\subsection{Zero-Energy States}

We will now analyze the 2D tunneling problem for the zero-energy electron-like states.
From the Hamiltonian in (\ref{eq:h1}) we can see that the momentum and mass of the particle can be non-zero even though the energy of the particle is zero. The eigenstates corresponding to the zero-energy electron/hole-like states are given by 
\begin{eqnarray}
u^\prime(k_x,k_y) = 
\left(
\begin{array}{cc}
\frac{m}{k_x+i k_y}\\
 0   \\
 0   \\
 1
\end{array}
\right),  \ \ \ \
%\end{eqnarray}
%
%\begin{eqnarray}
%
v^\prime(k_x,k_y) = 
\left(
\begin{array}{cc}
 0   \\
  \frac{m}{k_x-i k_y}   \\
1\\
0
\end{array}
\right)
\label{eigenstates:e0}
\end{eqnarray}
The wave functions in the three regions shown in Figure \ref{fig:2dbarrier} are as given in equations (\ref{eq:wave-fn}) with the respective eigenstates (\ref{eigenstates:e0}). Applying the continuity of the wave function yields the following transmission coefficients for the zero-energy states:
\begin{eqnarray}
t_{0^+}&=&\frac{4 e^{-i k_x d}k_x q_x}{(q_x+k_x)^2 e^{-i k_x d}-(q_x - k_x)^2 e^{i k_x d}} \\
\label{eq:transmission-e0}
%\end{eqnarray}
%
%\begin{eqnarray}
r_{0^+} &=& \frac{k_x-i k_y}{k_x+i k_y}\  \frac{ 2 i (k_x - q_x) (k_x + q_x) \sin(d q_x)}{(q_x+k_x)^2 e^{-i k_x d}-(q_x - k_x)^2 e^{i k_x d}}
\label{eq:rfcoeffs-e0}
\end{eqnarray}
with $|t_{0^+}|^2+|r_{0^+}|^2=1$. The coefficients corresponding to the hole-like states vanish , $t_{0^-}=r_{0^-}=0$. The transmission coefficient (\ref{eq:transmission-e0}) is the same as equation (\ref{eq:transmission}) and we find that $|r_E|^2+|r_{-E}|^2=|r_{0^+}|$. These coefficients are consistent with and describe the tunneling of non-chiral electrons across the barrier. These equations predict that a zero-energy electron cannot be transmitted or reflected as a zero-energy hole. Note that for $m=0$ in the Hamiltonian (\ref{eq:h1}), the transmission and reflection coefficients for the zero-energy state vanish, whereas those corresponding to the massless dispersion relation presented in the previous section are still non-zero.

\section{Conclusion} \label{conclude}

{We presented a {non-hermitian} Hamiltonian which can describe massless and zero-energy states in {graphene-like} condensed matter systems}. We solved the 2D scattering problem and derived the expression for the transmission and reflection of massless and zero energy electron/hole like states. Our analysis predicts that a massless electron-like state can be converted to a hole upon reflection from the barrier whereas it is always transmitted as an electron. For the zero-energy state however, transition to the hole-like state cannot take place upon reflection or transmission. Experimental observations are needed to test the predictions of this analysis.

\section{Acknowledgments}
The author would like to thank Mahtab A. Khan for useful discussions.

%\newpage

\end{document}